\documentclass[twocolumn]{article}
\usepackage{nolta2014}
\usepackage{txfonts}
\usepackage{graphicx}
\usepackage{cite}
\usepackage{multirow}
\usepackage{dcolumn}

\begin{document}

\title{Measuring Contextual Relationships in Temporal Social Networks by Circle Link}

\author{Cui Jing, Yiqing Zhang, and Xiang Li${}^\dag$}

\address{
Electronic Engineering Department\\
Fudan University\\
Shanghai, 200433, China\\
\dag Email: lix@fudan.edu.cn
}

\maketitle

\abstract
Network science has released its talents in social network analysis based on the information of static topologies. In reality social contacts are dynamic and evolve concurrently in time. Nowadays they can be recorded by ubiquitous information technologies, and generated into temporal social networks to provide new sights in social reality mining. Here, we define \emph{circle link} to measure contextual relationships in three empirical social temporal networks, and find that the tendency of friends having frequent continuous interactions with their common friend prefer to be close, which can be considered as the extension of Granovetter's hypothesis in temporal social networks. Finally, we present a heuristic method based on circle link to predict relationships and acquire acceptable results.

\endabstract
\section{Introduction}
In the past decades, we have witnessed fruitful and exciting advances in studies of complex, large-scale social networks. Many algorithms and methods have been brought up in predicting social relationships \cite{lu2011link} and modeling social network structure \cite{watts1998collective} based on local and global topological information of network. Since the developments of ubiquitous digital technologies in sensing, storage and communication, we have experienced the emergence of large-scale human behaviour data with high temporal resolution. Consequently, a new complex network concept, namely \emph{Temporal Networks}, has been presented\cite{holme2012temporal}; new methods using temporal information have been proposed for social network analysis, such as closeness recognition\cite{eagle2009inferring} and interactive patterns modeling\cite{perra2012activity}. A nascent interdisciplinary area, \emph{temporal social network}, is coming to the stage.

Inspired by the work of Song et al.\cite{song2010limits} about the predictability of human mobility, Takaguchi et al.\cite{takaguchi2011predictability} measured the predictability of social contacts in face-to-face networks applied the concepts of information entropy. They declare that prior knowledge of social contact between the ego and his (or her) current partner can decrease the uncertainty of social contact between the ego and the next partner by a large percentage. Followed by their conclusions, we define the contextual relationship between the current partner and the next partner as \emph{Circle Link}, characterizing the potential social circles. The empirical analyses of three temporal social networks show that there is no universal memory mechanism in human contextual relationships, and two vertices of dense edges have dense time-ordered edges with their common neighbor, which can be considered as the extension of Granovetter's hypothesis in temporal social networks. At the last part of this article, we give a heuristic method based on the definition of circle link for predicting relationships with limited, local prior knowledge of social contacts.

\section{Datasets}
Three datasets are used in our research, where two of them are collected by RFID technologies during the ACM Hypertext Conference in Torino and the \emph{INFECTIOUS: STAY AWAY} art-science exhibition at the Science Gallery in Dublin\cite{isella2011s}, respectively. The third one is derived from the Campus Wi-Fi login records in Fudan University\cite{zhang2013temporal}. We refer these three datasets as HT, SG and WF respectively, and use the data of three days in each dataset to generate the corresponding temporal social networks, whose detailed properties are shown in Table \ref{properties}.

\begin{table}[bth]
\renewcommand{\arraystretch}{1.3}
\caption{Properties of all temporal social networks}
\begin{center}
\begin{tabular}{c c|c c c c}
\hline
\multicolumn{2}{c|}{Dataset} & Nodes  & Records & Edges & Sparsity \\
\hline
\multirow{3}{*}{HT} & Day1  & 100    & 3460    & 946   & 5.93e-04  \\
                    & Day2  & 102    & 3510    & 1062  & 7.14e-04  \\
                    & Day3  & 97     & 2895    & 926   & 8.39e-04  \\
\hline
\multirow{3}{*}{SG} & Day1  & 200    & 2684    & 714   & 7.92e-04 \\
                    & Day2  & 204    & 2770    & 739   & 7.61e-04 \\
                    & Day3  & 186    & 2467    & 615   & 7.39e-04 \\
\hline
\multirow{3}{*}{WF} & Day1  & 1120   & 12833   & 10346 & 0.0120   \\
                    & Day2  & 2250   & 25772   & 21637 & 0.0067   \\
                    & Day3  & 1906   & 15798   & 13744 & 0.0057   \\
\hline
\end{tabular}
\end{center}
\label{properties}
\end{table}

\begin{figure}[thb]
\centering
\includegraphics[width=8cm]{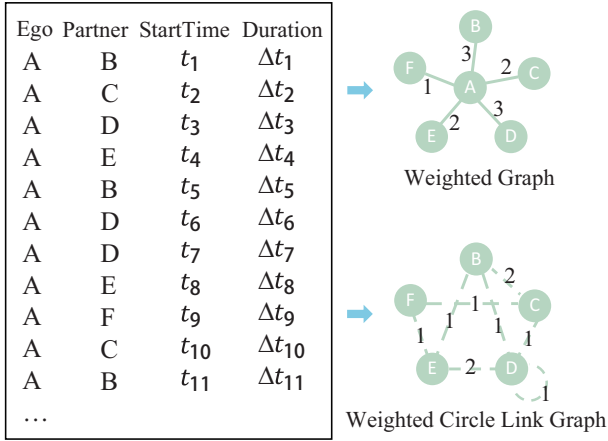}
\caption{The illustrations of Circle Link}
\label{clexample}
\end{figure}

Each network can be presented as a list of conversation events, consisting of two participants, the start time and the duration. To analyze the contextual relationships, we generate the event list of a specific ego as shown in the left panel of Fig.\ref{clexample}, ignoring the durations of events and being ordered by the start time. The corresponding (static) weighted social network is aggregated over the whole temporal social network, and the weight of edges is indicated by the number of events between two persons (see the right panel of Fig. \ref{clexample}).

\section{The Definition of Circle Link}

\begin{figure}[bht]
\centering
\includegraphics[width=8cm]{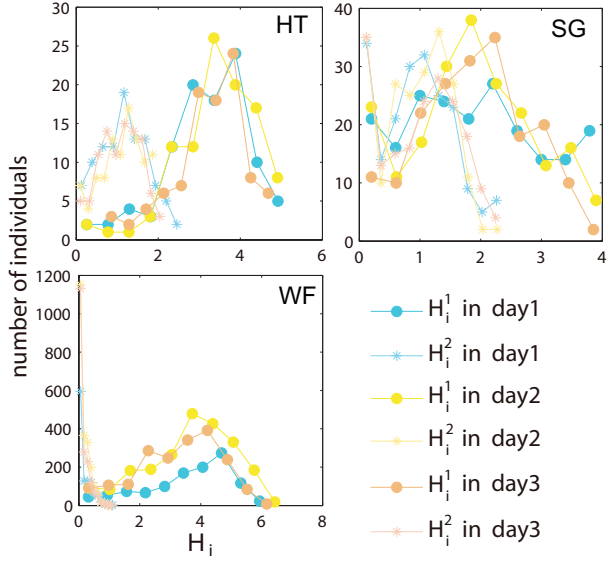}
\caption{Distribution of entropies in three datasets}
\label{entropy}
\end{figure}

We apply the definitions \cite{takaguchi2011predictability} of the uncorrelated entropy, ${H}_{i}^{1}=-\sum_{j\in{N}_{i}}{{P}_{i}(j)\log_{2}{{P}_{i}(j)}}$, and the conditional entropy,  ${H}_{i}^{2}=-\sum_{j\in{N}_{i}}{{P}_{i}(j)\sum_{l\in{N}_{i}}{{P}_{i}(l|j)\log_{2}{{P}_{i}(l|j)}}}$ in all temporal social networks (where ${N}_{i}$ is the set of ego \emph{i}'s partners/neighbors, ${P}_{i}(j)$ represents the historical probability that ego \emph{i} contacts partner \emph{j}, and ${P}_{i}(l|j)$ represents the conditional probability that ego \emph{i} contacts partner \emph{l} after a contact with partner \emph{j}). Fig. \ref{entropy} shows that the conditional entropies in the datasets of SG and WF are followed by normal distributions, in contrast to the results of \cite{takaguchi2011predictability}. Nevertheless, the conditional entropy is smaller than the corresponding uncorrelated entropy in all networks, indicating prior knowledge of event with the current partner can decrease the uncertainty of the event with the next partner. Therefore we define \emph{Circle Link} to capture the potential social relationship from such continuous contextual relationships. The weight of a circle link is defined as the cumulative number of such continuous contextual relationships and measures the potential degree. As shown in the left panel of Fig.\ref{clexample}, since the person $A$ has two continuous conversation events with his or her two partners $B$ and $C$ at the time $t_1,t_2$ and $t_{10},t_{11}$, the weight of the circle link between $B$ and $C$ indicated by $cl_{A}(BC)$ is $2$ (see the right panel of Fig.\ref{clexample}). We can define the mean weight of each circle link in network level as follows:
\begin{equation}
{ W }_{ CL }(i,j)\equiv \sum _{ l\in \mathrm{V},l\neq i,j }{ |{ cl }_{ l }(i,j)| }
\end{equation}
where $\mathrm{V}$ is the set of all nodes.

\section{Results}
\subsection{Self Circle Link Phenomenon}
The definition of circle link includes self-loop phenomenon. An individual can be circle linked to himself or herself, like individual \emph{D} in the right panel of Fig. \ref{clexample}. We define the following \emph{Self Circle Rate} (\textbf{SCR}) to represent the percentage of self circle links among all circle links observed by the ego:
\begin{equation}
{ \textbf{SCR} }_{ i }\equiv\frac { |{ cl }_{ i }(jj)| }{ |{ cl }_{ i }(lj)| } ,\quad l,j\in N_{ i },
\label{scr}
\end{equation}
where $N_i$ is the set of ego $i$'s partners.

We apply the null hypothesis that an ego contacts his or her partners without a memory mechanism, where $\textbf{SCR}^{null}_i=1/|N_{i}|$, and define the ratio $m_0$ between $\textbf{SCR}_{i}$ and $\textbf{SCR}_{i}^{null}$ averaged on the whole network as follows:
\begin{equation}
m_{ 0 }\quad \equiv \sum_{i\in \mathrm{V}} \frac{\textbf{SCR}_{i}}{\textbf{SCR}_{i}^{null}} =  \sum_{i\in \mathrm{V}} \frac{\textbf{SCR}_{i}}{1/N_i}
\label{mzero}
\end{equation}
where $\mathrm{V}$ is the set of all nodes. Table \ref{selfcl} shows that in the Hypertext Conference, contacts among people have memory mechanism ($m_{0}>1$ for HT), different from contacts in the Gallery setting ($m_{0}\approx1$ for SG). Moreover, in the human indoor interaction network people repel to contact with their current partner as the next one ($m_{0}\leq0.65$ for WF), suggesting that the model represented in \cite{karsai2012correlated} don't capture the mechanisms of all social contacts. Furthermore, $m_{0}$ is time invariant within a dataset, indicating the memory or inverse-memory mechanism is only determined by the contexts of social contacts.

\begin{table}[tbh]
\renewcommand{\arraystretch}{1.3}
\caption{The rate $m_{0}$ in temporal social networks}
\begin{center}
\begin{tabular}{c|c c c}
\hline
Dataset& Day1 & Day2 & Day3 \\
\hline
HT & 1.8794 & 1.9596 & 1.5307 \\
SG & 0.8308 & 0.7573 & 0.7497 \\
WF & 0.6517 & 0.5587 & 0.3470 \\
\hline
\end{tabular}
\end{center}
\label{selfcl}
\end{table}
\subsection{Strength and Clustering Coefficient of Social Ties Correlated with Circle Link Weights}

\begin{figure}[hbt]
\centering
\includegraphics[width=8cm]{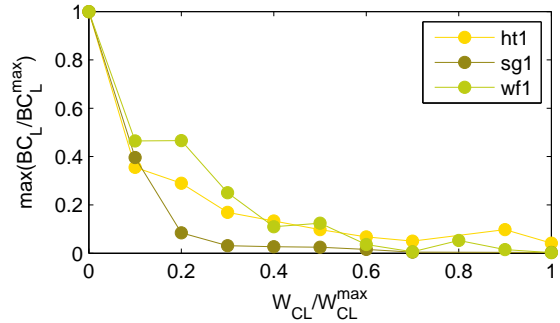}
\caption{Counter-relationship between $W_{CL}$ and $BC_{L}$}
\label{bc-wcl}
\end{figure}

In the aggregated version of temporal social networks, the weights of edges represent the strengths of social ties. Here we use the Pearson correlation coefficients $\rho$ to characterize the correlation between the strengths of social ties and temporal patterns. Table \ref{pea} shows that in all temporal social networks, the Pearson correlation coefficients $\rho_{W_L,W_{CL}}$ between the weight of edges $W_L$ and the weight of circle links $W_{CL}$ have relative high values $\rho_{W_L,W_{CL}}>0.5$, indicating that two vertices frequently continuously contact with their common neighbor have dense edges between them. Furthermore, Fig. \ref{bc-wcl} shows that the weights of circle links are inversely proportional to the corresponding link betweenness centralities, indicating that the involving vertices are in the dense local network. These are closely related to Granovetter's hypothesis that states that in social networks dense edges have on average higher weights \cite{granovetter1973strength}. However, in addition to having higher weights, we find that dense edges are more commonly related to ``continuous group talk'', temporal patterns involving three individuals.

Furthermore, we apply the definition of edge clustering coefficient \cite{pajevic2012organization} $CC_{L}(i,j) = { n }_{ C }(i,j) / { n }_{ T }(i,j)$ , where ${ n }_{ C }(i,j)$ is the number of common neighbors of individual \emph{i} and individual \emph{j}, ${ n }_{ T }(i,j)$ is the total number of nodes neighbored individual \emph{i} or individual \emph{j}. The high edge clustering coefficient represents dense overlaps of the corresponding two vertices' neighborhoods. As shown in Table \ref{pea}, the Pearson correlation coefficients $\rho_{CC_L,W_{CL}}$ between link clustering coefficients and weights of circle links have relative low values $\rho_{CC_L,W_{CL}}<0.5$ in two face-to-face networks, but relative high values $\rho_{CC_L,W_{CL}}>0.5$ in human indoor interaction network, indicating ``continuous group talk'' involving more than three individuals exits in human indoor interaction network, but not in two face-to-face networks.

\begin{table}[thb]
\renewcommand{\arraystretch}{1.3}
\caption{Pearson correlation coefficient of weight of circle links and other network metrics}
\begin{center}

\begin{tabular}{c c|c c}
\hline
\multicolumn{2}{c|}{}  & $\rho_{W_{L}, W_{CL}}$  & $\rho_{CC_{L}, W_{CL}}$    \\
\hline
 \multirow{3}{*}{HT} & Day1  & 0.7390    & 0.2851     \\
                     & Day2  & 0.7292    & 0.2382     \\
                     & Day3  & 0.5583    & 0.2248     \\
\hline
 \multirow{3}{*}{SG} & Day1  & 0.7031    & 0.4682     \\
                     & Day2  & 0.7170    & 0.3971     \\
                     & Day3  & 0.6993    & 0.4833     \\
\hline
 \multirow{3}{*}{WF} & Day1  & 0.5935    & 0.7532     \\
                     & Day2  & 0.5079    & 0.7745     \\
                     & Day3  & 0.5547    & 0.7701     \\
\hline
\end{tabular}

\end{center}
\label{pea}
\end{table}

\section{Relationship Prediction Method}
Here we present a heuristic method based on circle link to predict relationships in temporal social networks. We use the weights of circle links $W_{CL}$ as the predictor of potential relationships, where $W_{CL}>0$ represents there exits a potential relationship and vice verse. The weights of edges in weighted social networks $W_L$ are used to testify the predictor, where $W_L>0$ indicates there exits a relationship in real, and vice verse. Therefore, Table \ref{classify} shows four classifies of results in our prediction method. \emph{Precision} ($TP/(TP+FP)$) and \emph{recall} ($TP/(TP+FN)$) are used to quantify the exactness and completeness of our method.

\begin{table}[tbh]
\renewcommand{\arraystretch}{1.3}
\caption{The classifies of results in relationship prediction method}
\begin{center}
\begin{tabular}{c|c c}
\hline
   & $W_{L}>0$  & $W_{L}=0$ \\
\hline
$W_{CL}>0$ & True Positive(TP) & False Positive(FP) \\
$W_{CL}=0$ & False Negative(FN) & True Negative(TN) \\
\hline
\end{tabular}
\end{center}
\label{classify}
\end{table}
%
%
%
Table \ref{PR} shows that our method performs well in all temporal social networks, giving the evidence that it is possible to observer the structure of a large-scale social network by locating a few sensors and analyzing their temporal interaction data. The well performance of our method can be intuitively contributed to high positive correlation between the predictor $W_{CL}$ and the tester $W_L$. Moreover, high \emph{precision} is also caused by high positive correlation between the predictor $W_{CL}$ and the edge clustering coefficient $CC_{L}$ when comparing Table \ref{pea} with Table \ref{PR}. Another possible factor is the clustering coefficient of the network. It has been testified in our previous work \cite{cui2013clustering} that the average temporal clustering coefficient of WF is larger than those of face-to-face networks. We further calculated the following Circle Rate (\textbf{CR}) of the weighted and boolean clustering coefficient ($WCC$ and $CC$) of each node in the \emph{WF} dataset:
\begin{equation}
{ \textbf{CR} }_{ i }\equiv \frac { |{ cl }_{ l,j }\wedge l_{ l,j }| }{ |{ cl }_{ l.j }| } ,\quad l,j\in { N }_{ i }
\label{cr}
\end{equation}
T-test results as shown in Table \ref{ttest} give the evidence that high \emph{precision} is possibly caused by high clustering phenomenon in temporal social networks. Finally, the \emph{recall} of one face-to-face network (SG) is two times larger than that of another face-to-face network (HT), which is because that people are more temporally clustered in former network \cite{isella2011s}.
\begin{table}[tbh]
\renewcommand{\arraystretch}{1.3}
\caption{The precision and recall of our method in all temporal social networks*}
\begin{center}
\begin{tabular}{c c|c c}
\hline
\multicolumn{2}{c|}{Dataset}  & Precision  & Recall \\
\hline
\multirow{3}{*}{HT} & Day1  & 0.5023    & 0.2650 \\
                    & Day2  & 0.4781    & 0.2374 \\
                    & Day3  & 0.4413    & 0.2563 \\
\hline
\multirow{3}{*}{SG} & Day1  & 0.5322    & 0.5669 \\
                    & Day2  & 0.5501    & 0.5806 \\
                    & Day3  & 0.5681    & 0.6483 \\
\hline
\multirow{3}{*}{WF} & Day1  & 0.8484    & 0.2537 \\
                    & Day2  & 0.7824    & 0.2491 \\
                    & Day3  & 0.7788    & 0.2896 \\
\hline
\end{tabular}
\end{center}
*The precision and recall are averaged over all nodes.
\label{PR}
\end{table}

\begin{table}[bht]
\renewcommand{\arraystretch}{1.3}
\caption{T-test results about the hypothesis that the $\textbf{CR}_{i}$ has a larger mean than the $CC_{i}$ or the $WCC_{i}$ in \emph{WF} dataset*}
\begin{center}
\begin{tabular}{c|c c}
\hline
p-value & $Weighted CC$   & $Unweighted CC$ \\
\hline
Day1 & 2.7520e-07 & 2.7453e-10\\
Day2 & 0.0055     & 2.0769e-04\\
Day3 & 0.5311     & 0.1662\\
\hline
\end{tabular}
\end{center}
*The hypothesis is true when p-value is smaller than 0.005.
\label{ttest}
\end{table}
\section{Conclusions}
In this work, We defined a new term \emph{Circle Link} to measure the contextual relationship of ego and help predict potential relationship between ego's partners. The empirical analyses confirmed that the memory mechanism is not universal in all social contacts. Furthermore, the tendency of close friends having frequent continuous interaction with their common friend can be seen as an extension of Granovetter's hypothesis to temporal social networks. Finally, we presented a heuristic method of using contextual information to excavate potential relationship within ego's neighborhoods and discuss main influence factors. We believe future amelioration of this method would help to implement larger-scale data collection of temporal social networks with limited sensors.
\section*{Acknowledgments}
The authors acknowledged the SocioPatterns project for sharing their data on human face-to-face proximity contact, the Informatization Office of Fudan University for the WiFi Data collection. This work was partly supported by the National Key Basic Research and Development Program (No.2010CB731403), the NCET program (No.NCET-09-0317), and the National Natural Science Foundation (No.61273223) of China.


\end{document}